\documentclass[10pt,prd,aps]{revtex4}
\newcommand{\roothalf}{\mbox{$\textstyle \frac{1}{\sqrt{2}}$}}
\newcommand{\rootthird}{\mbox{$\textstyle \frac{1}{\sqrt{3}}$}}
\newcommand{\rootsixth}{\mbox{$\textstyle \frac{1}{\sqrt{6}}$}}
\newcommand{\threetwo}{\mbox{$\textstyle \frac{3}{2}$}}
\newcommand{\twothirds}{\mbox{$\textstyle \frac{2}{3}$}}

\begin{document}
%\preprint{}

\title[Unusual quantum states:]{Particle in infinite potential well with variable walls}
% {Particle in infinite potential well with variable boundary conditions} 
%Particle in infinite potential well with vanishing wall 
%{A counter-intuitive Quantum Mechanical Gedankenexperimemt}
%{Superluminal communication in non-relativistic Quantum Mechanics: A simple example}

\author{Bernhard K. Meister  }  
%\\\vspace{6pt}, Department of Physics, Renmin University of China, \\ Beijing, China 100872
%}
\email{bernhard.k.meister@gmail.com}
%\affiliation{...., Imperial College, London SW7
%2BZ, UK}
\affiliation{ Department of Physics, Renmin University of China, Beijing, China 100872}

\date{\today }

%\affiliation{Blackett Laboratory, Imperial College, London SW7
%2BZ, UK }

%\date{\today}

%\input{psfig.sty}

\begin{abstract}
\noindent  A Gedanken experiment is described to explore a counter-intuitive feature
%'property' in arxiv submission
 of quantum mechanics.
A particle is placed in a one-dimensional infinite well. The barrier on one side of the well is suddenly removed and the chamber   dramatically enlarged. %to a multiple of its original size.
At specific, periodically recurring,  times  the particle can be found with probability one at the opposite end of the enlarged chamber in an interval of the same size as the initial well.  With the help of symmetry considerations these times are calculated and shown to be dependent on the mass of the particle and the size of the enlarged chamber.  
Parameter ranges 
% for the mass of the particle and the size of the well pre- and post-enlargement
are given, where 
the non-relativistic nature of standard quantum mechanics becomes particularly apparent. %, if one chooses  judiciously. %, but not necessarily in a physically realisable way, .  

%{energy conservation; nonlocality; quantum
%measurement; Bell inequality; fractals; von Neumann entropy;
%Maxwell's daemon} 
\end{abstract}
\maketitle
%\vspace{-1.59cm}
%\section{Introduction}
%\label{s1}
  \vspace{-.3cm}
\noindent Quantum mechanics contains a tangle of unresolved foundational issues, involving  the role of probability, the measurement process, the reduction of the state vector, and the relation to  relativity, and provides  ample topics for investigation.
 In this note a simple Gedanken experiment 
%of limited practical applicability 
is considered to study an implication of the lack of upper limit for the speed of information propagation in the Schr\"odinger equation - a diffusion equation. %of information  of The Schroedinger equation is a diffusion equation with infinite speed of propagation, i.e. information propagates above any fixed speed limit.

A particle is put in a one-dimensional infinite well of size $\delta$. One of the barriers is removed at  time $t=0$ and the particle is able to spread out over a larger well of size $L$.   
 After a well-defined time, dependent  on the size of the  post-expansion well and the particle mass,  the wave function will again be concentrated in an interval of the initial size of the box, but this time at the other edge of the enlarged well.  The instantaneous well expansion has been extensively studied  by Aslangul\cite{asa} and in papers cited  therein.
%\hspace{.41cm}

 A particle of mass $m$,  trapped in a
one-dimensional infinite square well of width $\delta$,  is described by the Hamiltonian
\begin{eqnarray}
H=-\frac{\hbar^2}{2m}\frac{d^2}{d x^2} \label{eq:1.1}. \nonumber
\end{eqnarray}
The boundary conditions are such that the wave function $\psi(x)$
vanishes at $x=0$, $x=\delta$ and outside the well. 
The starting wave function $\psi(x)$ is chosen to be 
\begin{eqnarray}
\psi(x)=\sqrt{\frac{2}{\delta}}\sin\left(\frac{\pi  x}{\delta}\right)
\quad {\rm with}\quad E'=\frac{\pi^2\hbar^2}{2m\delta^2}.
\label{eq:1.2} \nonumber
\end{eqnarray}
For more details about quantum mechanics in an infinite well see the paper by Bender {\it et al.}\cite{bbm}.

After removing the barrier at $t=0$ the infinite potential is resized  and reaches from the origin to $L$ with $L>\delta$. The Hamiltonian stays unchanged, but the boundary conditions are replaced by 
the wave function vanishing at $x=0$ and $x=L$. %The new wave function has the form $\Psi(x,t)$.
The new stationary states and the
corresponding energy eigenvalues are
\begin{eqnarray}
\phi_n(x)=\sqrt{\frac{2}{L}}\sin\left(\frac{n\pi x}{L}\right)
\quad {\rm and}\quad E_n=\frac{\pi^2\hbar^2n^2}{2mL^2},
\label{eq:1.2} \nonumber
\end{eqnarray}
where $n$ is any whole number.
The new wave function in the enlarged well depends on the choice of the starting eigenfunction in the smaller well $\psi(x)$ and has the form 
\begin{eqnarray}
\Psi(x,t)=\sum_{n=1}^{\infty}a_{n} \phi_n (x)e^{\imath E_n t/\hbar }, \nonumber
\end{eqnarray}
with the  transition probability between the original test function and the new basis at the time the barrier is raised of the form
\begin{eqnarray}
a_{n}=  \frac{2}{\pi } \sqrt{\frac{\delta}{L}}
 \sin\Big(\frac{\pi n \delta }{L}\Big) \bigg(1-n^2\frac{\delta^2}{L^2}\bigg)^{-1}
= \frac{2}{\pi } \,\,
\frac{ \sin(\pi n \eta) }{1-n^2 \eta^2}\,\,\, \sqrt{\eta}\,\, \nonumber
%\frac{1}{ n^2 \pi}\frac{  \sin\Big(\frac{\pi n \delta }{L}\Big) }{  1-\frac{L^2 }{n^2 \delta^2}} \bigg(\frac{L}{\delta}\bigg)^{3/2}. \nonumber
\end{eqnarray}
with $\eta$ defined as $\delta/L$.
The probability of finding the particle in the interval $[L-\delta,L]$ at any time $t>0$   is 
\begin{eqnarray}
&&\int_{L-\delta}^{L} dx |\Psi(x,t)|^2 =\sum_{n=1}^{\infty}\sum_{m=1}^{\infty}\int_{L-\delta}^{L} dx \,\,a_{n} a_{m}^*  e^{\imath (E_n-E_m) t} \sqrt{\frac{2}{L}}\sin\left(\frac{n\pi x}{L}\right)\sqrt{\frac{2}{L}}\sin\left(\frac{m\pi x}{L}\right).\nonumber\\
%&=&\frac{4}{\pi^2}\sum_{n=1}^{\infty}\sum_{m=1}^{\infty} (-1)^{n+m}\frac{1}{n m}\cos((E_n-E_m) t)
%\frac{ \sin(\pi n \eta) }{\eta^2-1}
%\frac{ \sin(\pi m \eta) }{\eta^2-1}\eta
%\nonumber\\&&
%\Big(\frac{1}{ \pi (m-n)(m+n)}\Big)\Big((m+n) \sin\Big(\pi (\eta-1)(m-n) \Big)-(m-n) \sin\Big(\pi (\eta-1)(m+n)\Big)   \Big). \nonumber\\
%&=&\frac{4}{\pi^3}\sum_{n=1}^{\infty}\sum_{m=1}^{\infty} \frac{1}{n m}\cos((E_n-E_m) t)
%\frac{ \sin(\pi n \eta) }{\eta^2-1} 
%\frac{ \sin(\pi m \eta) }{\eta^2-1} 
%\nonumber\\&&
%\Big(\frac{\eta}{ (m-n)(m+n)}\Big)\Big((m-n) \sin\Big(\pi (m+n)\eta\Big) -(m+n) \sin\Big(\pi (m-n) \eta\Big)  \Big). \nonumber\\
%&=&-\frac{8}{\pi^3}\sum_{n=1}^{\infty}\sum_{m=1}^{\infty}\cos((E_n-E_m) t)
%\bigg( \frac{1}{ m}\frac{ \sin(\pi n \eta) }{\eta^2-1} 
%\frac{ \cos(\pi m \eta) }{\eta^2-1}-  \frac{1}{n }\frac{ \cos(\pi n \eta) }{\eta^2-1} 
%\frac{ \sin(\pi m \eta) }{\eta^2-1}\bigg)
%\Big(\frac{\eta}{ (m-n)(m+n)}\Big). \nonumber
%&=&\frac{2}{L}\sum_{n=1}^{\infty}\sum_{m=1}^{\infty}  \cos(E_n-E_m) t)\frac{  \sin\Big(\frac{\pi n b }{L}\Big)}{\pi L^2- \pi b^2 n^2}b L^2\frac{  \sin\Big(\frac{\pi m b }{L}\Big)}{\pi L^2- \pi b^2 m^2}b L^2
\end{eqnarray}
Instead of evaluating the  sum, we  shall calculate  the probability at specific times, where symmetry properties come into play. 
The simplest case  is to choose $\hat{t}$ in such a way that $\exp(\imath E_n \hat{t}/\hbar)$ %=\exp\Big(\imath \frac{\pi^2\hbar^2n^2}{2mL^2} \hat{t}\Big)$ 
has for even and odd   $n$  always the values
$+1$ and $-1$, respectively:
\begin{eqnarray}
{\rm for\,\,\,\,n=2k}:
\exp\Big(\frac{\pi^2\hbar(2k)^2}{2mL^2} \hat{t}\Big)=  \exp(2\pi\imath ) =1\nonumber 
\end{eqnarray}
and
\begin{eqnarray}
{\rm for\,\,\,\,n=2k+1}:  \,\,\,\,\exp\Big(\frac{\pi^2\hbar(2k+1)^2}{2mL^2} \hat{t}\Big)=  \exp( \pi\imath) =-1,\nonumber 
\end{eqnarray}
where $k$ is any natural number.
The smallest possible solution is
\begin{eqnarray}
%\rightarrow
\hat{t}:= \frac{2mL^2}{\pi\hbar},\nonumber 
\end{eqnarray}
others can be generated by multiplying the original result with any odd whole number. The time $\hat{t}$  could be called a  `revival time',  since the reflected wave function in its original form is reconstituted in its entirety at a new time and place. %For the higher order starting eigenstates of the form $j$ the critical time $\hat{t}_n$ scales like $j^{-2}$ and is equal to $2mL^2/(\pi\hbar j^{2})$.
At  time $\hat{t}$ the wave function is concentrated solely in the interval $[L-\delta,L]$, since it corresponds, except for a overall phase,  exactly to the mirror-symmetric starting wave-function at time $t=0$ in the interval $[0,\delta]$. In the rest of the box, corresponding to the interval $[0,L-\delta]$, the wave function is zero. A more detailed explanation is given next. Odd and even elements  of the $\sin (n \pi x/L)$  basis functions have different behaviour at the beginning and the end of the interval. The even basis functions change sign at the two extreme ends of the interval, i.e. $\sin (2k \pi x/L)=-\sin (2k \pi (L-x)/L)$, whereas all the odd basis functions have the same sign, i.e. 
 $\sin ((2k+1) \pi x/L)=\sin ( (2k+1)\pi (L-x)/L)$. At time $\hat{t}$  the energy pre-factor $\exp(\imath E_n \hat{t}/\hbar)$  toggles between plus and minus one for even and odd terms, which compensates the sign change of the basis elements, resulting in %the set up at $t=0$ on the left edge is reproduced on the right edge at $\hat{t}$, i.e. 
$|\psi(x)|^2=|\Psi(x,0)|^2= |\Psi(L-x,\hat{t})|^2$. 

At  time  $\hat{t}$  the particle has reconstituted itself in the $\delta$-interval at the far end of box. If at this time   light has not reached the other end of the well, i.e. $  L/c \gg \hat{t} $, and $\delta$ is small compared to $L$,  then this implies a contradiction between relativity with its limit on the velocity of signals and the prediction of quantum mechanics. 
%can be used  to construct a Gedankenexperiment to test the compatibility of linear Schroedinger equation based quantum mechanics with  the restriction to supra-luminal signals.
 The inequality above can be rewritten as $\hbar  \pi \gg 2 m  L c$. 
One might be interested to inquire, if there are parameter ranges for which the inequality is satisfied. %The subsequent estimation of possible parameter values is preliminary and will possibly be extended in a different setting.
As an example, one might select for the particle mass and the two length scales involved
 \begin{eqnarray}
 \hat{\delta}\sim &10^{-15}\,\,{\rm m} \,\,\,\,\,  &{\rm the \,\,\,`size` \,\,\,of \,\,an \,\,\,electron},\nonumber\\ 
 \hat{L}\sim &10^{-13}\,\,{\rm m} \,\,\,\,\,  &{\rm 100 \,\,\,\,\,times \,\,\,the \,\,\,`size` \,\,\,of \,\,an \,\,\,electron},\nonumber\\ 
 \hat{m}\sim& \,\,\,9\times 10^{-31} \,\,{\rm kg} \,\,\,\,\,\,& {\rm mass\,\,\,of \,\,\,an \,\,\,electron}. \nonumber
 \end{eqnarray}
The %first microscopic distance $\hat{L}$ provides a time 
time $\hat{t}$ in this particular case is of the order of $10^{-22}s$, while the time required for light to transverse the distance   $\hat{L}$ is around $10^{-13} m/ (3 \times 10^{8} m/s)\sim  10^{-21} s$.  %In addition $\hat{\delta}$ is assumed to be around $10^{-15}{\rm m}$. 
This suggests a natural barrier at which quantum mechanics   scrapes  against the relativistic limit, i.e. fails to comply with restrictions placed by relativity, since $\hat{t}$ is smaller by a factor of roughly ten than the time it takes light to transverse across $\hat{L}$. %A simple-to-state  experiment could be used to check, if the particle is found on the far side of the well at the time specified. 
One could contemplate carrying out such an experiment to confirm the location of the particle at the time specified. Naturally, this is simpler to propose than to realise. 

In this note a limiting case was explored to understand the superluminal aspect of non-relativistic quantum mechanics. 
 This 
was studied with the help of a Gedanken experiment and is possibly instructive, since the notion that super-luminal communication has been banished from non-relativistic quantum mechanics is  assumed by some in a cavalier fashion.

Idealisations have been liberally employed. Three in particular spring to mind: First, the removal of the barrier to allow the expansion of the wave function onto the whole length of the interval
happens instantaneously. Compressing an extended process into a point in time,    like the occurrence of an instantaneous  measurement,  is not an unusual calculation tool in quantum mechanics. %Any instantaneous process is %How does one construct a solvable time-dependent Hamiltonian to mimic this behaviour?
The case of a moving wall has been analysed extensively, e.g. see the recent paper by Cooney \cite{cooney}, which includes a review of different approaches going back to the work  by Hill and Wheeler in the 1950s.  A challenge is to find solutions, where the energies stay real. 
 The removal of a wall differs from moving a wall with finite velocity, but similarities exist.    Second,  the energy of the particle post barrier removal  does not differ  from the initial energy, but the energy variance changes dramatically. %e.g. a similar question was considered in \cite{bbm} and in some papers by the author. 
 How does the situation change, if the barrier removal is handled in a more realistic way? % and suggest that the barrier removal is associated with a huge inflow of energy into the system. 
Third, the choice of parameters %, i.e. particle mass and the length of the initial and final potential well,  
to meet the constraints of the inequality above are not necessarily experimentally feasible.
% While being cognisant of these caveats, the note shows there are box sizes and particle masses, where  standard quantum mechanics breaks down unceremoniously in a Gedankenexperiment setting. 
 
 The approach described has some similarities  to  Maxwell's fishpond \cite{kisler} from classical physics. In both proposals  an excitation, e.g. pebble thrown in a pond,  is reconstituted  after a well-defined time in a specific, but  separate, place. %Due to its one dimensional nature, maybe the name  quantum  Maxwell canal might be considered apt.  %since  the wave function can be reconstituted at the arrival interval at well-defined time.
 
 This elicits the question, what applications one could devise, besides checking the validity of quantum mechanics. One could use such a  device to  transport  besides information also energy. % or to
 %provide a form of teleportation. % in a different and  more restricted form, since  a signal at the origin can be broken into a superposition of eigenfunctions. Subsets of these eigenfunctions have different transportation times, i.e. $\hat{t}$, and 
 % in distinct chunks.  
 %After one batch arrives the barrier is lowered at $L-\delta$ and the particular part transferred to a waiting area, where it is then with the arrival of the last batch recombined to  form the original wave function.
  A series of these boxes applied in sequence could be used to amplify the effect.  Experimental relevance will be considered  separately, since the emphasis of this note is on simplicity not applicability.  
  
  The paper by Aslangul\cite{asa} gives a   detailed analysis of the  instantaneous well expansion, while this note tries to point out a potential implication for  quantum mechanics.
 
While being  cognisant of the  various idealisations employed, the   Gedanken experiment  presented  puts  bounds,  dependent on the post-expansion size of the well and the mass of the test particle, on the applicability of non-relativistic quantum mechanics, if super-luminal communication is to be avoided.\\
 \noindent  Helpful comments  by L.P. Hughston  and an email  by Prof. Aslangul drawing  my attention to his  paper\cite{asa} are gratefully acknowledged. 
%\end{acknowledgements}


\vspace{-.162cm}
\begin{enumerate}
\bibitem{asa} C. Aslangul, {\it J. Phys. A} {\bf 41}, 075301 (2008).


  \bibitem{bbm} C. M. Bender, D.C.  Brody\& B.K. Meister, % Unusual quantum states: nonlocality, entropy, Maxwell's daemon, and fractals.
        {\it Proceedings of the Royal Society London} {\bf A461}, 733-753 (2005), ~arXiv:quant-ph/0309119.

\bibitem{cooney} K. Cooney, ~arXiv:1703.05282.

 \bibitem{kisler} P. Kinsler {\it et al.}, {\it Eur. J. Phys.} {\bf 33} 1737
       (2012), ~arXiv:1206.0003.

               
\end{enumerate}
\end{document}